# Identifying Semantically Duplicate Questions Using Data Science Approach: A Quora Case Study

Navedanjum Ansari
*navedanjum.ansari@gmail.com*

Rajesh Sharma
*rajesh.sharma@ut.ee*

## ABSTRACT

Identifying semantically identical questions on, Question and Answering(Q&A) social media platforms like Quora is exceptionally significant to ensure that the quality and the quantity of content are presented to users, based on the intent of the question and thus enriching overall user experience. Detecting duplicate questions is a challenging problem because natural language is very expressive, and a unique intent can be conveyed using different words, phrases, and sentence structuring. Machine learning and deep learning methods are known to have accomplished superior results over traditional natural language processing techniques in identifying similar texts.

In this paper, taking Quora for our case study, we explored and applied different machine learning and deep learning techniques on the task of identifying duplicate questions on Quora's question pair dataset. By using feature engineering, feature importance techniques, and experimenting with seven selected machine learning classifiers, we demonstrated that our models outperformed previous studies on this task. Xgboost model with character level term frequency and inverse term frequency is our best machine learning model that has also outperformed a few of the Deep learning baseline models.

We applied deep learning techniques to model four different deep neural networks of multiple layers consisting of Glove embeddings, Long Short Term Memory, Convolution, Max pooling, Dense, Batch Normalization, Activation functions, and model merge. Our deep learning models achieved better accuracy than machine learning models. Three out of four proposed architectures outperformed the accuracy from previous machine learning and deep learning research work, two out of four models outperformed accuracy from previous deep learning study on Quora's question pair dataset, and our best model achieved accuracy of 85.82% which is close to Quora state of the art accuracy.

Authors addresses: Navedanjum Ansari, Institute of Computer Science, University of Tartu, 50090 Estonia; Rajesh Sharma, Institute of Computer Science, University of Tartu, 50090 Estonia.

## KEYWORDS
Quora, Duplicate question, Machine learning, Deep learning, model, neural network

## 1 INTRODUCTION

Social media platforms are a great success as can be witnessed by the number of the active user base. In the age of internet and social media, there has been a plethora of social media platforms, for example, we have Facebook, for user interaction, LinkedIn, for professional networking, WhatsApp for chat and video calling, Stack Overflow for technical queries, Instagram for photo sharing. Along the line, Quora is a Question & Answer platform and builds around a community of users to share knowledge and express their, opinion and expertise on a variety of topics.

Question Answering sites like Yahoo and Google Answers existed over a decade however they failed to keep up the content value [32] of their topics and answers due to a lot of junk information posted; thus their user base declined. On the other hand, Quora is an emerging site for the quality content, launched in 2009 and as of 2019, it is estimated to have 300 million active users[1]. Quora has 400,000 unique topics[2] and domain experts as its user so that the users get the first-hand information from the experts in the field.

With the growing repository of the knowledge base, there is a need for Quora to preserve the trust of the users, maintain the content quality, by discarding the junk, duplicate and insincere information. Quora has successfully overcome this challenge by organizing the data effectively by using modern data science approach to eliminate question duplication.

### 1.1 Research Problem

As for any Q&A, it has become imperative to organize the content in a specific way to appeal users to be an active participant by posting questions and share their knowledge in respective domain of expertise. In keeping the users' interest, it is also essential that users do not post duplicate questions and thus multiple answers for a semantically similar question, this is avoided if semantically duplicate questions are merged then all the answers are made available under the same subject. Detecting semantically duplicate questions and finding the probability of matching also helps the Q&A platform to recommend questions to the user instead of

---
[1]Vox - https://www.vox.com/recode/2019/5/16/18627157/quora-value-billion-question-answer
[2]Statistics 2019 - https://foundationinc.co/lab/quora-statistics/

posting a new one. Given our focus of study, we defined the following two research questions:

**RQ1:** How can we detect duplicate questions on Quora using machine learning and deep learning methods?

**RQ2:** How can we achieve the best possible prediction results on detecting semantically similar questions ?

Research questions one and two have been studied on the first dataset released by Quora[3], however our aim is to achieve the higher accuracy on this task.

## 1.2 This Work

We have extracted different features from the existing question pair dataset and applied various machine learning techniques. After employing feature engineering upon raw dataset, we experimented with different machine learning algorithms to draw our baseline. We also showed that not all features were useful in predicting duplicate question and after analyzing and dropping a few of the features, our result for ML models slightly improved but did not degrade at all. We also have the existing baseline from the works of literature, which we have surpassed. We then tried many deep learning methods to finally experiment with our four best deep learning architectures. With our experiment results, we have shown that deep learning methods are suitable for solving the problem of detecting semantically similar questions. Our deep learning neural networks performed better than baselines from previous research studies.

Moreover, our machine learning ensemble model TF-IDF achieved the accuracy of 82.33% and higher F1 score compared to literature [31]. Also, our best deep learning model achieved an accuracy of 85.82%. Three out of four presented deep learning models outperformed the results from the literature [1, 6, 30, 31] and our best result achieved close to Quora's state of the art accuracy presented by Quora engineering team on their blogpost [20].

## 2 LITERATURE REVIEW

The previous work to detect duplicate question pairs using Deep learning approach [1], shows that deep learning approach achieved superior performance than traditional NLP approach. They used deep learning methods like convolutional neural network(CNN), long term short term memory networks (LSTMs), and a hybrid model of CNN and LSTM layers. Their best model is LSTM network that achieved accuracy of 81.07% and F1 score of 75.7%. They used GloVe word vector of 200 dimensions trained using 27 billion Twitter words in their experiments.

The method proposed in [17] makes use of Siamese GRU neural network to encode each sentence and apply different distance measurements to the sentence vector output of the neural network. Their approach involves a few necessary steps. The first step was data processing, which involves tokenizing the sentences in the entire dataset using the Stanford Tokenizer[4] . This step also involved changing each question to a fixed length for allowing batch computation using matrix operations. The second step involves sentence encoding, where they used both recurrent neural network(RNN) and gated recurrent unit (GRU). They initialized the word embedding to the 300-dimensional GloVe vectors [27].

The next step was determining the distance measure [21] that are used in combining the sentence vectors to determine if they are semantically equivalent. There were two approaches for this step, the first being calculating distances between the sentence vectors and running logistic regression to make the prediction. The paper has tested cosine distance, Euclidean distance, and weighted Manhattan distance. The problem here is that it is difficult to know the natural distance measure encoded by the neural network. To tackle this issue, they replaced the distance function with a neural network, leaving it up to this neural network to learn the correct distance function. They provided a row concatenated vector as input to the neural network and also experimented using one layer and two- layer in the neural network. The paper utilized data augmentation as an approach to reduce overfitting. They also did a hyperparameter search by tuning the size of the neural network hidden layer (to 250) and the standardized length of the input sentences (to 30 words) which led to better performance.

In the literature [30], authors have used word ordering and word alignment using a long-short-term-memory(LSTM) recurrent neural network [10], and the decomposable attention model respectively and tried to combine them into the LSTM attention model to achieve their best accuracy of 81.4%. Their approach involved implementing various models proposed by various papers produced to determine sentence entailment on the SNLI dataset. Some of these models are Bag of words model, RNN with GRU and LSTM cell, LSTM with attention, Decomposable attention model.

LSTM attention model performed well in classifying sentences with words tangentially related. However, in cases were words in the sentences have a different order; the decomposable attention model [26] achieves better performance. This paper [26] tried to combine the GRU/LSTM model with the decomposable attention model to gain from the advantage of both and come up with better models with better accuracy like LSTM with Word by Word Attention, and LSTM with Two Way Word by Word Attention.

In the relevant literature [31], the authors have experimented with six traditional machine learning classifiers. They used a simple approach to extract six simple features such as word counts, common words, and term frequencies(TF-IDF) [28] on question pairs to train their models. The best accuracy reported in this work is 72.2% and 71.9% obtained from binary classifiers random forest and KNN, respectively.

Finally, we reviewed the experiments by Quora's engineering team [20]. In production, they use the traditional machine learning approach using random forest with tens of manually extracted features. Three architectures presented in their work use LSTM in combination with attention, angle, and distances. The point noted from this literature is that Quora uses the word embedding from its Quora Corpus whereas all other selected baselines from the literature review used GloVe [27] pre-trained word to vectors from the glove project[5].

---

[3]https://www.quora.com/q/quoradata/First-Quora-Dataset-Release-Question-Pairs
[4]https://nlp.stanford.edu/software/tokenizer.shtml
[5]https://nlp.stanford.edu/projects/glove

Table 1: Performance Baseline from selected literature

| Paper | Model | Technique | Acc | F1 score |
|---|---|---|---|---|
| Detection of Duplicates in Quora and Twitter Corpus [31] | Logistic Regression | Machine Learning | 0.671 | 0.66 |
| | Decision Tree | | 0.693 | 0.69 |
| | KNN | | 0.719 | 0.72 |
| | Random Forest | | **0.722** | **0.73** |
| Determining Entailment of Questions in the Quora Dataset [30] | LSTM | Deep learning | 0.784 | 0.8339 |
| | LSTM with Attention | | 0.81 | 0.8516 |
| | LSTM with Two Way Word by Word Attention | | **0.814** | **0.8523** |
| | Decomposable Attention Model | | 0.798 | 0.8365 |
| Quora Question Duplication [6] | Siamese with bag of words | Deep learning | 77.3 | 73.2 |
| | Siamese with LSTM | | 83.2 | 79.3 |
| | Seq2Seq LSTM with Attention | | 80.8 | 76.4 |
| | Ensemble | | **83.8** | **79.5** |
| Duplicate Question Pair Detection with Deep Learning [1] | LSTM (twitter word embedding 200d) | Deep learning | **0.8107** | **0.757** |
| Quora State of the Art [20] | LSTM with concatenation | Deep learning | 0.87 | 0.87 |
| | LSTM with distance and angle | | 0.87 | 0.88 |
| | Decomposable attention | | 0.86 | 0.87 |

The results achieved in each of the previous studies on Quora duplicate question pair dataset is summarized as presented in Table 1

## 3 DATASET

In this section, we briefly describe the data collection, exploratory data analysis, data visualization, and data cleaning process.

### 3.1 Data collection

The data for this research work is taken from the First Quora Dataset release hosted on Amazon S3[6]. There is a total of 404290 rows in the dataset, which indicates that there are total 404290 question pairs, and the overall file size is 55.4 MB.

GloVe pre-trained word vectors are used for word embeddings. GloVe [27] pre-trained vectors are available at SNLI project site Glove. To convert word to vector for distance calculation, we used Google news vectors [25]*GoogleNews-vectors-negative300.bin.gz*, of 3 million words and 300 dimensions.

### 3.2 Data Exploration

We performed the necessary statistics on the dataset, which helps us to give a more detailed understanding of the duplicate Quora question dataset. There is a total of six columns in the dataset. Each of the columns is meaningful and describe the characteristic of the row. The description of the columns is as described below in Table 2.

### 3.3 Dataset Representation

Table 3 contains the total number of question pairs and the distribution of class labels. Positive samples are those identified as semantically duplicates and negative samples are non-duplicate pairs.

[6]http://qim.fs.quoracdn.net/quora_duplicate_questions.tsv

Table 2: Description of columns in dataset

| Colum Name | Description |
|---|---|
| id | A unique identifier assigned to each row in the dataset. The first row has an id of 0, and the last row has id 404289 |
| qid1 | A unique identifier for the question in question1 column. |
| qid2 | A unique identifier for the question in question2 column. |
| question1 | question1 contains the actual question to be compare d with question2 |
| question2 | question2 contains the actual question to be compare d with question2 |
| is_duplicate | is_duplicate is the result of a semantical comparison of question pair. 0 indicates false i.e. question pair is not duplicate 1 indicates true i.e. question pair is duplicate |

Table 3: Class Label distribution

| | |
|---|---|
| Positive Sample (1) | 149263 |
| Negative Sample (0) | 255027 |
| Total Question Pairs | 404290 |

In the histogram plot Figure 1, the x-axis represents the number of times question occurs, and the y-axis or height of the bar represents how many such questions with occurrence count exist in the dataset. As can be visualized from the graph the majority of questions occurs less than 60 times, and the first bar shows the unique occurrence and second bar the number of the appearance of question twice and so on.

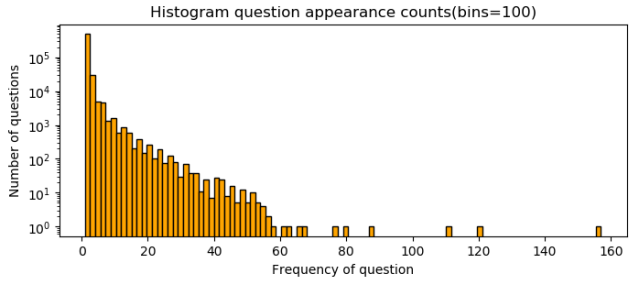

Figure 1: Distribution of question occurrence in dataset

### 3.4 Data Cleaning

We computed, additional stats on our dataset that helps us explore the data and make decision in eliminating redundant data rows.

Table 4: Statistics on question1 and question2

| Statistics | Average | Sum | Count |
|---|---|---|---|
| q1 length | 59.53672 | 24070099 | 404290 |
| q2 length | 60.10838 | 24301217 | 404290 |
| Max length(char count) q1 | | | 623 |
| Max length(char count) q2 | | | 1169 |
| q1 length - q2 length | -0.57166 | -231118 | 404290 |
| q1 length <=5 | - | - | 53 |
| q2 length <=5 | - | - | 19 |

Mostly these questions short length questions are one word, one and two length questions are just the question marks and special characters, foreign characters. We discard as these data rows in the data cleaning process. In Table 4 we can see that the q2 length on an average is greater, and therefore, we have an average negative difference. We dropped a total of 72 rows from our raw dataset based on the logic that both question1 length and question2 less than 6 or either one of the question length is less than six.

Thus, we have 404218 data rows in our machine learning experiments, and we continue with the usual data with 404290 rows for our deep learning experiments.

## 4 BACKGROUND

This section briefly explains the features extracted from the raw dataset and various machine learning and deep learning neural layers used in the experiments.

### 4.1 Feature Engineering

We dropped the first three columns id, qid1, and qid2 from the initial raw dataset and created additional useful features so that we have two columns question1, question2, and class label *is_duplicate* and 28 new derived features, Therefore initially, we have total of thirty-one columns in dataset provided as input to the machine learning classifiers.

**Set 1 Original Feature**

1. **Question 1 dataset:** This is the question1, column in the dataset.

2. **Question 2 dataset:** This is the question2, column in the dataset.

3. **Is duplicate:** Class label represented as *1* for duplicates and *0* for non-duplicates.

**Set 2 Basic Features**

4. **Length of question1:** Length of the question1, includes all the characters, punctuation and white spaces.

5. **Length of question 2:** Length of the question2, includes all the characters, punctuation and white spaces.

6. **Difference in the length of questions:** Difference between the length of corresponding question1 and question2.

7. **Number of characters in q1:** Distinct number of characters excluding white spaces in corresponding question1.

8. **Number of characters in q2:** Distinct number of characters excluding white spaces in corresponding question2.

9. **Number of words in q1:** Number of words in question1 including repeated words.

10. **Number of words in q2:** Number of words in question2 including repeated words.

11. **Number of common words in q1 and q2:** Distinct common words in corresponding question1 and question2.

**Set 3 Fuzzy Feature**

12. **Qratio:** Qratio feature is the quick ratio comparison of the two question strings and has value range from 0 to 100. More similar questions have a higher score.

13. **Wratio:** Wratio feature is the weighted ratio that uses different algorithms to calculate the matching score and returns the best ratio for two question strings. Score range from 0 to 100.

14. **Partial ratio:** Partial ratio feature calculates the best score for partial string matching against all sub strings of the greater length and returns the best score. Score range from 0 to 100.

15. **Token set ratio:** Token set ratio [33] feature is calculated on the strings by segregating the strings into three parts. First part of common strings which are then arranged as sorted intersection, and other parts from each of the questions as sorted remainders. It then computes scores from compares sorted intersection with each of combination of sorted intersection and sorted remainders of that string. Score range from 0 to 100.

16. **Token sort ratio:** Token sort feature tokenizes the strings and then sort the strings alphabetically and join back into strings. It then compares the transformed strings using ratio to return score. Score range from 0 to 100.

17. **Partial token set ratio:** Partial token set feature is similar to token set ratio except that after it tokenizes string it uses partial ratio in place of ratio to calculate the matching score. Score range from 0 to 100.

18. **Partial token sort ratio:** Partial token sort ratio is similar to token sort ratio except that it uses partial ratio in place of ratio, after sorting the token to compute matching score. Score range from 0 to 100.

**Set 4 Distance Features**

**19. Word mover's distance(wmd):** Word mover's distance [23] feature calculates the distance between two documents, in our case, it gives the distance between two corresponding questions in our dataset. It uses word2vec embedding to find the distance between similar or semantically similar words. The stop words like ' *the*,' '*to*' etc. are removed using nltk [2] library.

**20. Normalized word mover's distance (norm wmd ):** Normalized word mover's is similar to word mover's distance just that word2vec vectors are normalized, normalizing helps in reducing risk of incorrect computation.

**21. Cosine distance:** Cosine distance feature calculates the angle between the word vectors of two question sentences.

**22. Minkowski distance:** Minkowski distance feature is a generic distance metric that can be computed as the summation of differences of vector dimensions raise to the power p and whole raise to the inverse of power p. We have used p=3 to calculate the Minkowski distance.

**23. Cityblock distance:** Cityblock distance feature is a special case of Minkowski distance metric when we use the value of p=1 in the equation of Minkowski distance.

**24. Euclidean distance:** Euclidean distance feature is also a special case of Minkowski distance metric when we use the value of p=2 in the equation of Minkowski distance.

**25. Jaccard distance:** Jaccard distance [7] feature is computed as a ratio of intersection between two vectors sets to the union of two vector sets. The two vector sets are derived from the two question sentences in our dataset.

**26. Canberra distance :** Canberra distance is computed as the sum of the absolute difference of two vector points divided by the absolute sum of individual vector points.

**27. Braycurtis distance:** Braycurtis distance [34] is also called as Sorenson distance. It is also a variant of Manhattan distance normalized by the sum of the vector points in two objects x and y.

**Set 5 Vectors Features**

**28. Skew question1 vector:** Skewness is the measure of distribution [24] . Skewness indicates a deviation tendency from the mean in one of the direction. Skewness is computed over question1 vector. A normal distribution has a skew value equal to 0.

**29. Skew question2 vector:** Skewness is computed over question 1 vector.

**30. Kurtosis question1 vector:** Kurtosis distance is the measure of dense distribution towards the tails of the distribution [24]. A normal distribution has a value equal to 0. Kurtosis vector is computed over question1 vector.

**31. Kurtosis question2 vector:** Kurtosis is computed over question2 vectors.

## 4.2 Machine Learning Models

We have selected the following seven machine learning classifiers and a statistical feature TF-IDF.

**K-Nearest neighbors:** The k-nearest neighbors (KNN) [13] algorithm is a simple, easy-to-implement supervised machine learning algorithm that can be used to solve both classification and regression problems.

**Decision Tree:** Decision tree [29] is the most powerful and accessible tool for classification and prediction.

**Random forest:** Decision trees are the building blocks of the random forest model. Random forest [16], like its name implies, consists of a large number of individual decision trees that operate as an ensemble.

**Extra Trees:** Extra tree [11] classifier is a type of ensemble learning technique which aggregates the results of multiple uncorrelated decision trees collected in a " forest " to output its classification result.

**Adaboost:** AdaBoost [8] is a popular boosting technique which helps you combine multiple " weak classifiers " into a single " strong classifier ". A weak classifier is simply a classifier that performs poorly but performs better than random guessing.

**Gradient Boosting Machine:** Gradient boosting [9] is a machine learning technique for regression and classification problems, which produces a prediction model in the form of an ensemble of weak prediction models, typically decision trees.

**XGBoost:** XGBoost [3] is an implementation of gradient boosted decision trees designed for speed and performance. XGBoost is a decision-tree-based ensemble Machine Learning algorithm that uses a gradient boosting framework. XGBoost is short for extreme gradient boosting.

**TF-IDF** : TF-IDF [28] stands for term frequency -inverse document frequency, is a scoring measure widely used in information retrieval (IR). TF-IDF is intended to reflect how relevant a term is in a given document.

## 4.3 Elements of Neural Network Layers

**1. LSTM** [10]: Long short-term memory (LSTM) is an artificial recurrent neural network (RNN) architecture used in the field of deep learning. Unlike standard feed forward neural networks, LSTM has feedback connections. It can process not only single data but also entire sequences of data. LSTM networks are well-suited to classifying, processing, and making predictions based on time series data since there can be lags of unknown duration between essential events in a time series.

**2. Word Embedding** [23] : Word embeddings are a family of natural language processing techniques aiming at mapping semantic meaning into a geometric space. This is done by associating a numeric vector to every word in a dictionary, such that the distance between any two vectors would capture part of the semantic relationship between the two associated words.

**3. Glove Embedding** [27] : GloVe is used for obtaining vector representations for words. Training is performed on aggregated global word-word co-occurrence statistics from a corpus, and the resulting representations showcase interesting linear substructures of the word vector space.

**4. Time Distributed(Dense)** : Time distributed dense layer is used on RNN, including LSTM, to keep one-to-one relations on input and output. Assume we have 60 - time steps with 100 samples of data (60 x 100 in another word) and you want to use Recurrent Neural Network(RNN) with the output of 200. If we do not use time distributed dense layer, we will get 100 x 60 x 200 tensors. So we have the output flattened with each time step mixed.

**5. Lambda:** Lambda layer is a layer that wraps an arbitrary expression. For example, at a point, we want to calculate the square of a variable, but we can not only put the expression into our model because it only accepts layer, so we need Lambda function to make our expression be a valid layer in keras.

**6. Convolution 1D :** A CNN works well for identifying simple patterns within our data that will then be used to form more complex patterns within higher layers. A 1D CNN is handy when we expect to derive interesting features from shorter but mostly fixed-length segments of the overall data set and where the location of the feature within the segment is not of high relevance.

**7. GlobalMaxPooling 1D** [12] : This block performs precise ly the same operation as the 1D Max pooling block except that the pool size is the size of the entire input of the block, i.e., it computes a single max value for all the incoming data. The 1D Global max pooling block takes a vector and computes the max value of all values for each of the input channels. The output is thus a tensor of size is 1 x 1 x (input channels). Using 1D Global max pooling block can replace the fully connected blocks of our CNN

**8. Merge** [5] : Merge is used to join multiple neural networks together. A good example would be where we have two types of input, for example, tags and an image To combine these networks into one prediction and train them together, w e merge these Dense layers before the final classification.

**9. Dense** [18] : A dense layer is just a regular layer of neurons in a neural network. Each neuron receives input from all the neurons in the previous layer, thus densely connected. The layer has a weight matrix W, a bias vector b, and the activations of previous layer a.

**10. Batch Normalization** [19] : Batch normalization is a technique for improving the performance and stability of neural networks, and also makes more sophisticated deep learning architectures work in practice. The idea is to normalize the inputs of each layer in such a way that they have a mean output activation of zero and standard deviation of one. This is comparable to how the inputs to networks are standardized. How does this help? We know that normalizing the inputs to a network helps it learns. However, a network is just a series of layers, where the output of one layer becomes the input to the next. That means we can think of any layer in a neural network as the first layer of a smaller subsequent network. Thought of as a series of neural networks feeding into each other, we normalize the output of one layer before applying the activation function, and then feed it into the following layer (sub-network).

**11. Dropout** [15] : Dropout is a regularization technique, which aims to reduce the complexity of the model to prevent overfitting. Using "dropout," we randomly deactivate specific units (neurons) in a layer with a certain probability p from a Bernoulli distribution. So, if we set half of the activations of a layer to zero, the neural network w ill no t be able to rely on particular activations in a given feed-forward pass during training. As a consequence, the neural network will learn different, redundant representations; the network can no t rely on the particular neurons and the combination (or interaction) of these to be present. Another good side effect is that the training will be faster. Dropout is a technique used to tackle overfitting.

**12. PreLU** [14] : Parametric Rectified Linear Unit(PreLU), Parametric ReLU is inspired by ReLU, which, as mentioned before, has a negligible impact on accuracy compared to ReLU. Based on the same ideas that of ReLU, PreLU has the same goals: increase the learning speed by not deactivating some neurons. The primary argument for Parametric ReLu's over standard ReLu 's is that they do not saturate as we approach the ramp. In most other ways, they do not offer a distinct advantage. Think of it as an advantage in being able to tell the difference between a wrong answer and a horrible answer. The effect may not seem dramatic, but in some instances, it can be genuinely advantageous.

**13. Activation** [22] : Applies an activation function to the output of a layer such as *tanh, sigmoid* activation. It takes into consideration the effects of different parameter interaction and applies the transformation where it filters the value from which neuron to be passed to the next layer or the output.

## 5 METHODOLOGY

In this section, a general approach to training our machine learning classifiers, the process flow for feature importance analysis, the process of TF-IDF with ML classifiers and four different deep learning architectures that we modeled for our experiments are presented.

### 5.1 Experimental and Research Design

Influenced by the literature and the previous study, we started our experiments with the binary classification of whether a given pair of question is a semantically duplicate question. We began with feature engineering to produce as many as 28 new features from the given question pair dataset and apply different machine learning classifiers.

### 5.2 Feature Importance

We analyzed and studied the features extracted using feature engineering to validate the positive contributions from each of the features, and then we retrain our models by dropping the least important features. We have a total of 28 new features extracted in the experiment stage of section 4.1. We analyze and select the top twenty features that are helpful to our machine learning classifiers, and then dropped eight features.

### 5.3 Machine Learning Pipeline with TF-IDF

**TF-IDF character level**

The flow of TF-IDF character level feature with machine learning classifiers are presented inFigure 2. TF-IDF character level as the name suggests computes TF-IDF at character level in the document, in our case, it is a question.

The model learns the inverse frequency of characters from the set of combined unique question1 and question2 character set. The corresponding TF-IDF, character level feature, obtained for each of the questions in the pair is then passed as input to the different machine learning classifiers. The classifiers are then trained on the training dataset, which is 80% of total dataset and tested on 20% of the validation set. We also experimented with word level TF-IDF in a similar way as character level.

### 5.4 Deep Neural Network Design

**Architecture-1** : In this simple neural network architecture, we use a pair of questions as the two inputs. The architecture consists of the Embedding layer, LSTM layer applied separately on each of the question inputs, and then the model is merged using the Merge layer from keras library [4]. The output from the merged model layer is then passed through the series of Batch Normalization, Dense, Parametric rectified linear unit, Dropout and Sigmoid Activation function is applied at the final output layer. Embedding layers is the first hidden layer of a network that uses word embedding to represent a word as a dense vector, and we specify three arguments to the Embedding function, the input dimension, output dimension, and the input length. We use the input length, i.e. number of words as 40 and output dimension as 300. Input dimension is computed as the index of words + 1 in the sequence.

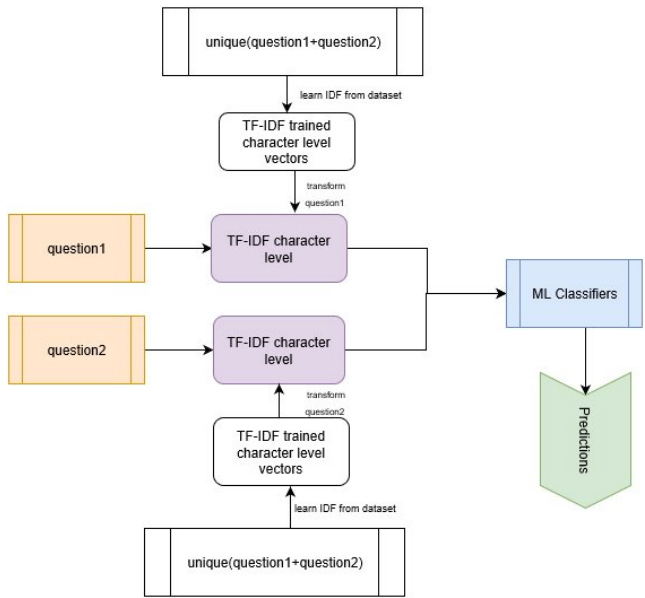

**Figure 2: The flow of TFIDF character level feature as an input to machine learning classifiers**

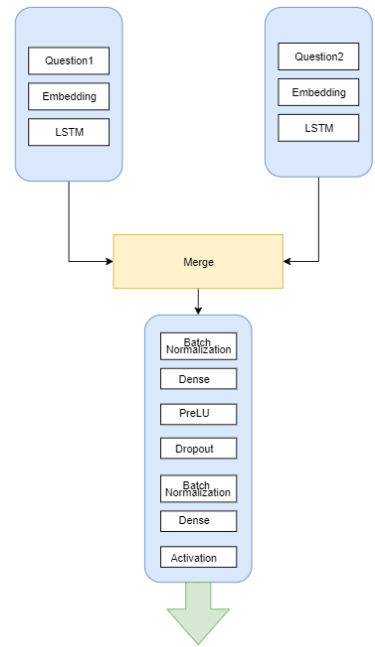

**Figure 3: Architecture-1 Simple Neural network architecture with two inputs**

In this model, we are not using any special pre-trained vectors like GloVe. The output of Embedding layer is fed to the LSTM layer. We used the dropout weight of 0.2 within LSTM to avoid overfitting. Each of the models merged as passed through a sequence of layers, as shown in Figure 3 The output from the intermediate Dense layer is 300, and the final Dense layer always has output dimension one, which then fed to sigmoid activation to give us the classification result.

**Architecture-2:** Neural network architecture-2 is modeled slightly different before applying to merge of different models otherwise after merge it very similar and trained on exactly same hyper-parameters as simple neural network presented as in Figure 3. In Architecture-2, we increase the number of independent models before merge to four, which are then merged and trained to produce the classification result. Architecture-2 with four inputs, two different networks are used for each of the questions as can be seen in Figure 4

Additional models before the merge, consist of Embedding layer using GloVe pre-trained vector of 300 dimension s with 840B tokens. Embeddings are then fed to Time distributed dense layer to maintain one to one relationship over time- distribution. Lambda sum is applied along the axis to produce the output of 300 dimensions. Thus, all the four independent models producing the output of 300d are then merged and passed through hidden layers of Batch Normalization, Dense, PreLu, Dropout, Batch Normalization, Dense and Sigmoid Activation to produce the classification result.

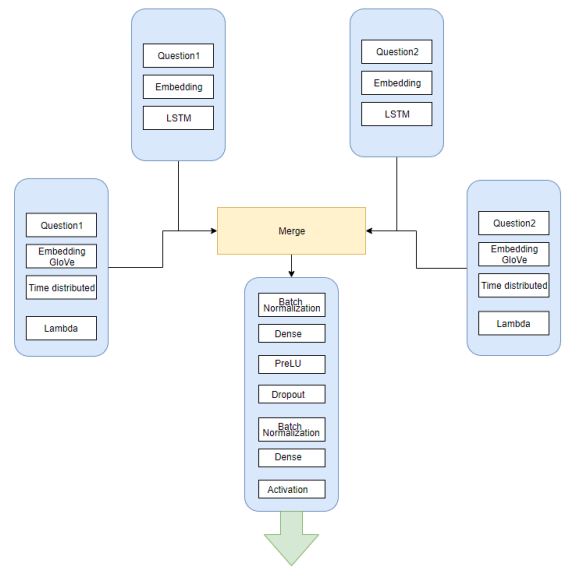

**Figure 4: Architecture-2 Deep neural network architecture with four inputs**

**Architecture-3** : Architecture-3 uses four sub-model or independent model from Architecture-2 with all the hyper-parameters tuned with the exact same value; the model differs after the merge of the four independent models. The modeled neural network architecture-3 can be visualized, as presented in Figure 5

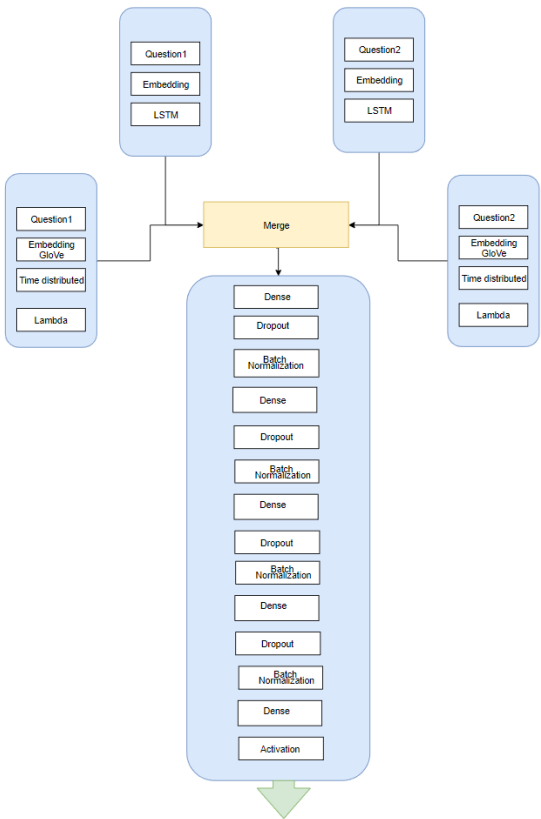

**Figure 5: Architecture-3 Deep neural network with four inputs and dense hidden layers**

**Architecture 4** : The deep neural network architecture-4 is modeled in such a way that it takes the six input which are then passed through six independent models and then merged into a single model consisting of twenty-three layers.

The deep neural network architecture-4 is modeled in such a way that it takes the six input which are then passed through six independent models and then merged into a single model consisting of twenty-three layers.

Four out of six independent or sub-models are similar to that of the four sub- models before the merge as presented in Figure 12. The two new sub-models that we added consist of GloVe based Embedding layer, Convolution Neural Network layer applied multiple times before and after Dropout layer. The output from the Convolution 1D layer is maxed out using Global Max Pooling 1D layer. Global Max Pooling output is then passed through hidden layers of Batch Normalization, Dense and Dropout. The Dropout layer has shown to perform well within our experiments with a weight of 0.2; therefore, throughout our neural network modeling ; dropout weigh used is 0.2. All six layers produce the output of dimension 300 which is then merged as a single model and passed through another twenty-six layer consisting of repeated units of Dense, Dropout and Batch Normalization and finally a Dense layer with the output of dimension size one which is fed to sigmoid Activation to predict the classification result. We have used tensorFlow keras [4] python library to model each of the neural network architecture presented in this section. All models are trained on the batch size of 300 and number of epoch iterations as 150.

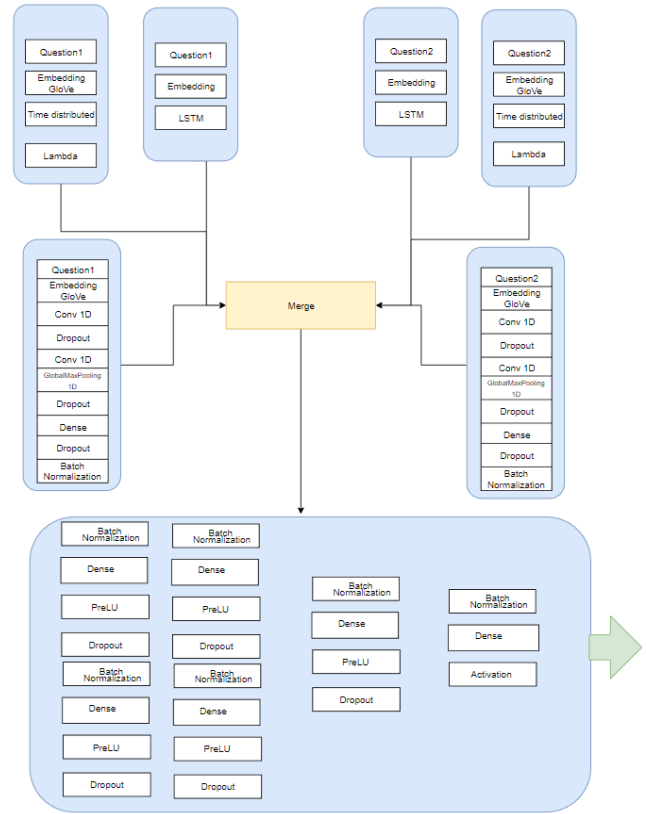

**Figure 6: Architecture-4 Deep neural network with six inputs and dense hidden layers**

## 6 DESCRIPTION OF MODELS AND RESULTS EVALUATION

This section discusses evaluation metrics and comparative analysis of the results.

## 6.1 Evaluation Metrics

The selection of metrics is the most crucial step in the evaluation of our models as it influences how we measure the performance of our model against each other and the baselines.

**Accuracy:** Accuracy is the ratio of the total number of correct predictions made by the models to the total number of predictions requested to the model.

**F1-Score:** F1-score or F1-measure is harmonic mean of precision and recall. To understand F1-Score, we need to understand Precision, also known as Specificity and Recall, also known as Sensitivity.

**Precision:** Precision or Specificity is the ratio of predicted positive samples that are actually positive to the total number of positive predictions made by the models.

**Recall:** Recall or Sensitivity is the ratio of predicted positive samples that are actually positive to the total number of actual positive predictions in total sample.

**Log loss:** Log loss is also known as cross-entropy, and when the classification type is of binary as in our research, then it is known as binary cross-entropy. Log Loss value lies in the range of {0,1} where ideal models will have log loss of 0, and the worst model will have log loss of 1. Log loss indicates how badly our model predicted the probability of our classification.

## 6.2 Baseline Model Classifiers

We trained our model and then evaluated the prediction on our test data set to achieve the baseline for our machine learning algorithms used in this research. Table 5 shows test accuracy and F1 score of our baseline machine learning models.

**Table 5: The baseline performance of traditional machine learning classifiers on the dataset with 30 features predicted on test dataset**

| Classifiers | Acc | F1-Score |
|---|---|---|
| K Nearest Neighbors | 0.7275 | 0.7031 |
| AdaBoost | 0.7041 | 0.6936 |
| XGBoost | 0.7417 | 0.7326 |
| Gradient Boost | 0.7271 | 0.7176 |
| Decision Tree | 0.7054 | 0.6992 |
| Random Forest | 0.7099 | 0.7016 |
| ExtraTrees | 0.7039 | 0.6849 |

As can be observed from Table 5, clearly the Xgboost model outperforms all the other selected classifiers with the Accuracy of 0.7416 and F1 score of 0.7326.

## 6.3 Feature Importance Analysis

We analyzed the feature importance value of all seven machine learning classifiers used in the experiments and executed the experiments. Based on our feature importance values, we selected the top 20 features out of 28 derived features . The performance result achieved after feature importance analysis and feature drop is as presented in Table 6

Xgboost, Gbm and KNN after feature drop still stood to be the top three performers in our baseline model set, and none of the classifiers suffers from any degradation. However, the gain achieved after feature drop is minimal. Figures 7 and 8 show the comparative visualization of Accuracy and F1 score before and after the feature drop. The eight dropped features are *difference in the length, WRatio, jaccard distance, braycurtis distance, Euclidean distance, cityblock distance, partial token set ratio, partial token sort ratio*.

**Table 6: Performance of traditional machine learning classifiers after feature drop on test dataset**

| Classifiers | Acc | F1-Score |
|---|---|---|
| K Nearest Neighbors | 0.7311 | 0.7076 |
| AdaBoost | 0.7048 | 0.6938 |
| XGBoost | **0.7431** | **0.7349** |
| Gradient Boost | 0.7289 | 0.7196 |
| Decision Tree | 0.7054 | 0.6992 |
| Random Forest | 0.7085 | 0.7021 |
| ExtraTrees | 0.7069 | 0.6914 |

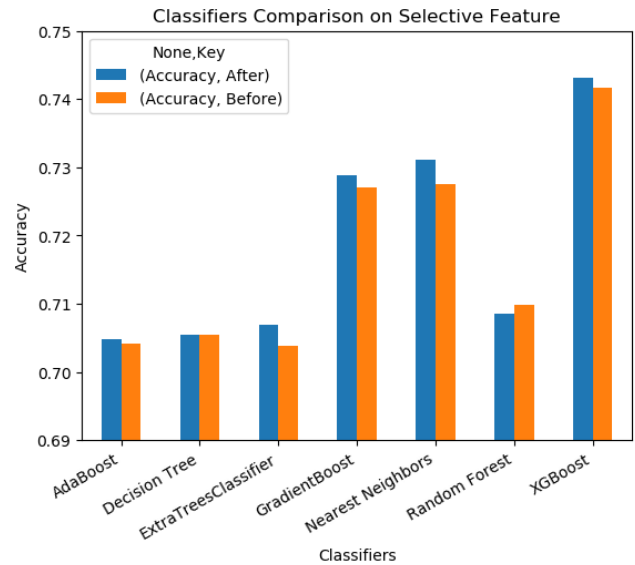

Figure 7: Accuracy comparison of ML classifiers Before versus After feature drop

## 6.4 TF-IDF with ML Models

Xgboost algorithm achieved an F1 score of 80.44 % compared to F1 score 79% published in [30] The accuracy achieved is 82.44%, which is very close to that of 83.7% achieved by the same literature. Thus, our result s show that ML models like Xgboost can also produce effective results similar to the Deep learning algorithms like LSTM.

Presented in Table 8, training and test accuracy and log loss metrics obtained from the deep neural network architectures presented in Figures 3, 4, 5 and 6.

Since we modeled and experimented with applied deep learning techniques using tensorflow keras library which offers only accuracy as the metrics at the end of each epoch and finding additional metrics like F1 score require us to run additional tests on test dataset and, calculate other metrics from prediction results.

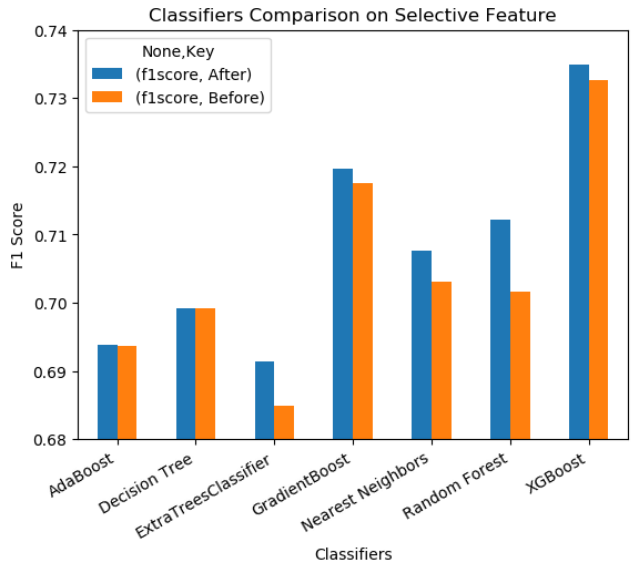

**Figure 8: F1 score comparison of ML classifiers Before versus After feature drop**

**Table 7: Performance of ML classifiers with TF-IDF word and TF-IDF character level on test dataset**

| Classifiers | Word TF-IDF | | Char TF-IDF | |
|---|---|---|---|---|
| | Acc | F1-Score | Acc | F1-Score |
| K Nearest Neighbors | 0.7513 | 0.7359 | 0.7845 | 0.7543 |
| AdaBoost | 0.6883 | 0.6076 | 0.6871 | 0.6201 |
| XGBoost | 0.7881 | 0.7596 | **0.8244** | **0.8044** |
| Gradient Boost | 0.6756 | 0.5339 | 0.6951 | 0.6009 |
| Decision Tree | 0.6677 | 0.5651 | 0.6672 | 0.5767 |
| Random Forest | 0.6284 | 0.3866 | 0.6484 | 0.4066 |
| ExtraTrees | 0.6281 | 0.3864 | 0.6581 | 0.4059 |

**Table 8: Accuracy and Log loss performance of deep neural network architectures evaluated on 20% of test dataset**

| Network | Train Loss | Train Acc | Test Loss | Test Acc |
|---|---|---|---|---|
| Architecture-1 | 0.2902 | 0.8715 | 0.4062 | 0.8133 |
| Architecture-2 | 0.2502 | 0.9012 | 0.4172 | 0.8312 |
| Architecture-3 | 0.1728 | 0.9127 | 0.4393 | 0.8522 |
| Architecture-4 | 0.0997 | 0.9674 | 0.38501 | 0.8582 |

## 7 CONCLUSIONS AND FUTURE WORK

We ensure that, the train and test data is split into 80/20 respectively throughout the experiments. We also ensure that the class labels in the test data set has proportionate distribution of samples as in our original dataset. All the hyper-parameters are selected based on grid search performed on the 10% of dataset from the training set, thus we ensure that our result do not suffer from overfitting.

Our results with TF-IDF and ML classifiers show that not all models performed well in ensemble with TF-IDF character level, but our best model Xgboost achieved the accuracy of 82.44 % and F1 score of 0.8044. This has demonstrated that machine learning models are efficient in solving natural language problem of detecting semantically similar question and compared to other baseline achieved from few of the deep learning methods such as LSTM and LSTM with Siamese listed in Table 1, our machine learning TF-IDF with Xgboost outperformed them.

Finally, we experimented with many different deep network layers and chose the four architecture to present which outperformed the results obtained by literature [6], our best performance from architecture-4 has achieved accuracy of 85.82 %. We used log loss measures for our neural networks along with accuracy. We reached the best training accuracy of 96.74% and log loss of 0.09 ; however, in our work, the test accuracy and test loss is our main focus. We achieved a better result from the previous study on the duplicate question pair dataset. Our best performance from this research work is the accuracy of **85.82% and log loss of 0.385**.

Our accuracy result is very near to the Quora state of the art [20] accuracy of 87%. The main reason for difference in results exist because Quora has used their own word embedding's from the Quora corpus dataset which is very specific to the Quora's question format, etc. whereas we have used the GloVe general embedding ; thus our results are methods are more relevant to any general question and answering system.

Another way, Quora could achieve a better by pre-processing the original question pair dataset. Since knowing the context in which question is asked, a proper replacement of some of the pronouns can be done, and higher accuracy can be achieved. For example, pronoun like us, we, they can be replaced if the topic under which question exists thus replacing it with their relative context like "American," " Programmers " and " Prisoners ' etc. during the pre-processing data stage can help achieve a better result. As we are unaware in which context questions were asked we could not do such pre-processing on the original dataset.

The limitations expressed in the paragraph above, if known in any context in case of any other social Media platforms or Quora can be used as the future development of this research. As also we worked on standard Intel core seven laptop, 32 GB RAM, without additional GPU capacity it took over 700 hours approx. to train all our four deep learning models and also the TF-IDF+Xgboost model training process took close to 7 hours. With better GPU capacity, we assume to achieve a slightly better result, and the experiment could have been performed with constructing more deep learning models and hyper parameter tuning.